\documentclass[aps,preprint,showpacs,nofootinbib,12pt]{revtex4}
\usepackage{graphicx}% Include figure files
\usepackage{epsfig}
\usepackage[dvips]{color}

\begin{document}

\title{Vertex functions for d-wave mesons in the light-front approach}

\author{ Hong-Wei Ke$^{1}$   \footnote{khw020056@hotmail.com} and
         Xue-Qian Li $^{2}$  \footnote{lixq@nankai.edu.cn}
       }

\affiliation{
  $^{1}$ School of Science, Tianjin University, Tianjin 300072, China \\
  $^{2}$ School of Physics, Nankai University, Tianjin 300071, China
  }

\begin{abstract}
\noindent While the light-front quark model (LFQM) is employed to
calculate hadronic transition matrix elements, the vertex
functions must be pre-determined. In this work we derive the
vertex functions for all d-wave states in this model. Especially,
since both of $^3D_1$ and $^3S_1$ are  $1^{--}$ mesons,  the
Lorentz structures of their vertex functions are the same. Thus
when one needs to study the processes where $^3D_1$ is involved,
all the corresponding formulas for $^3S_1$ states can be directly
applied, only the coefficient of the vertex function should be
replaced by that for $^3D_1$. The results would be useful for
studying the newly observed resonances which are supposed to be
d-wave mesons and furthermore the possible 2S-1D mixing in $\psi'$
with the LFQM.

\end{abstract}


\maketitle

\section{introduction}
In the hadronic physics  the most tough job is to calculate the
hadronic transition matrix elements which are fully governed by
non-perturbative QCD. Because of lack of solid knowledge on
non-perturbative QCD so far, one needs to invoke phenomenological
models. Applications of such models to various processes have
achieved relative successes so far. Among these models the
light-front quark model(LFQM) is a relativistic model and has
obvious advantages for dealing with the hadronic transitions where
light hadrons are involved \cite{Terentev:1976jk,Chung:1988mu}.
The light-front wave function is manifestly Lorentz invariant and
expressed in terms of fractions of internal momenta of the
constituents which are independent of the total hadron momentum.
This approach has been applied to many processes and thoroughly
discussed in literatures
\cite{Jaus:1999zv,Jaus:1989au,Ji:1992yf,Cheng:1996if,Cheng:2003sm,Hwang:2006cua,Ke:2009ed,Ke:2009mn,Li:2010bb,Wei:2009nc,Choi:2007se}.
Generally the results qualitatively coincide with experimental
observation and while taking  the error ranges into account (both
experimental and theoretical), they can be considered to
quantitatively agree with data.

However earlier researches with the LFQM  only concern the s-wave
and p-wave mesons whereas the higher orbital excited states have not
been discussed yet. With the improvements of experimental facilities
and rapid growth of database many new resonances have been observed
and some of them are regarded as higher orbital excited states, for
example $\psi(3770)$ and $\psi(4153)$ are suggested to be d-wave
charmonia whose principal quantum numbers are respectively $n=1$ and
$n=2$\cite{Lange:2010dp}. New analysis on $X(3872)$ indicates that
it has a possible charmonium assignment $^1D_2$\cite{Lange:2010dp}.
Moreover, a d-wave state $\Upsilon(1^3D_2)$ was observed by
CLEO\cite{Bonvicini:2004yj}. The situation persuades us to extend
our scope to involve d-wave. {It is generally believed that only the
lattice theory indeed deals with the non-perturbative QCD effects
from the first principle. So far,  the lattice study is constrained
by not only the computing abilities, but also the theory itself.
Even so, remarkable progresses have been made on the two aspects. It
is hoped that the lattice calculation will eventually solve all the
problems on hadrons, such as the hadron spectra, wavefunctions, even
the hadronic transition matrix elements. At present, the lattice
calculation indeed shed some light on the
wavefunctions\cite{Kroger:2009ek,Abada:2001yt,Arriola:2010up}. For
examples, the authors of Ref.\cite{Kroger:2009ek} suggested a method
to compute the spectra and wavefunctions of hadron excited states
and applied their method to the $U(1)_{2+1}$ lattice gauge theory.
Abada $et.\, al.$\cite{Abada:2001yt} study the pion light-cone wave
function on the lattice by considering the three-point Green
functions. On other aspects, for example, the readers who are
interested in the unquenched lattice calculation on the charmonia
whose total quantum numbers $J^{PC}$ are determined simultaneously,
are suggested to refer to those enlightening papers
\cite{Dudek:2008sz}. Some works about the radiative decays of
charmonia  are included in concerned
references\cite{Dudek:2009kk,Chen:2011kp}. More similar works can be
found in most recent
literatures\cite{Hagler:2011zz,Bazavov:2009ii,Dudek:2006ut}. In
fact, due to the rapid progress of lattice calculations, people are
more tempted to trust those results, but it by no means implies that
we should abandon phenomenological models because those models are
directly invented to manifest the physical mechanisms and moreover,
they are simpler and applicable in practice. }

To evaluate the transition rate in the LFQM one needs to know the
wave functions of parent and daughter hadrons. For any
$^{2S+1}L_J$ state, its wavefunction is constructed as the
corresponding spinors multiplying the so-called vertex function
which should be theoretically derived. It is also noted the
wavefunctions for s-wave and p-wave have been derived and their
explicit forms are given in Ref.\cite{Cheng:2003sm}. Following the
same strategy we obtain the wavefunction for all d-wave states.

The traditional LFQM was employed to study the decay constants and
form factors of weak decays
\cite{Jaus:1999zv,Ji:1992yf,Cheng:1996if}, but to maintain the
Lorentz covariance other contributions such as
Z-diagram\cite{Cheng:1996if} or
zero-mode\cite{Jaus:2002sv,Choi:2011xm,Choi:2010be,Cheng:1997au}
contributions must be included.

Thus a covariant LFQM\cite{Jaus:1999zv} has been suggested which
systematically includes the zero-mode contributions. In the
traditional LFQM the constituent quarks in the bound state are
required to be on their mass shells, nevertheless in the covariant LFQM
approach the constituent quarks in the meson are off-shell while
only the meson is on its mass shell. In this approach, one  writes
down the transition amplitudes  where all quantities and the
integrations maintain their four-dimensional forms, then integrates out
the light-front momentum $p^-$ in a proper way. While
carrying out the contour integration the antiquark is enforced to
be on its mass shell. The integrand in the remaining
three-dimension integration reduces into a form where all
quantities can be expressed in terms of the conventional
wavefunctions. During this procedure, some extra contributions
emerge comparing with the original scheme (see the text for
details).

In this work
after this introduction we derive the phenomenological vertex functions
for d-wave in the conventional light-front approach in section
II. Then in section III we  present their forms in the covariant
light-front approach.
In section IV we  discuss some formula for $^3D_1$
states and the section V is devoted to a brief summary.

\section{Vertex functions in the conventional light-front approach}
Let us first derive the vertex functions in the conventional light-front
approach.

In the conventional light-front approach a meson with the
total momentum $P$ and spin $J$ can be written as \cite{Cheng:2003sm}
\begin{eqnarray}\label{vf1}
|M(P,\,^{2S+1}L_{J},J_{Z})\rangle&&=\int\{d^3\tilde{p}_1\}\{d^3\tilde{p}_2\}2(2\pi)^3\delta^3(\tilde{P}-\tilde{p_1}-\tilde{p_2})\nonumber\\&&
\sum_{\lambda_1\lambda_2}\Psi^{JJ_z}_{LS}(\tilde{p_1},\tilde{p_2},\lambda_1,\lambda_2)|q_1(p_1,\lambda_1)\bar{q_2}(p_2,\lambda_2)\rangle,
\end{eqnarray}
where the flavor and color indices are omitted; $q_1$ and
$\bar{q_2}$ correspond to the quark and antiquark  in the meson
and $p_1$ and $p_2$ are the on-shell light-front momenta of quark
and antiquark, $\mathbf{p}$ is the three-momentum
$(\mathbf{p}_1-\mathbf{p}_2)/2$ and we define
\begin{eqnarray*}\label{vf2}
&&\tilde{p_i}=(p_i^+,p_{i\perp}),\,p_{i\perp}=(p_i^1,p_i^2),\,p_i^-=\frac{m^2+p_{i\perp}^2}{p_i^+},\,\{d^3\tilde{p_i}\}\equiv\frac{dp_i^+d^2p_{i\perp}}{2(2\pi)^3},\,\,
\\&&|q_1(p_1,\lambda_1)\bar{q_2}(p_2,\lambda_2)\rangle=b^\dagger_{\lambda_1}(p_1)d^\dagger_{\lambda_2}(p_2)|0\rangle.
\end{eqnarray*}
The light-front momenta $p_1$ and $p_2$ are expressed via the variables $p$ and $x_i\; (i=1,2)$ as
\begin{eqnarray}\label{vf3}
&&p_1^+=x_1P^+,\,p_2^+=x_2P^+,\,x_1+x_2=1,\nonumber\\&&
p_{1\perp}=x_1P_\perp+p_\perp,\,p_{2\perp}=x_2P_\perp-p_\perp.
\end{eqnarray}
\begin{center}
\begin{figure}[htb]
\begin{tabular}{c}
\scalebox{1.0}{\includegraphics{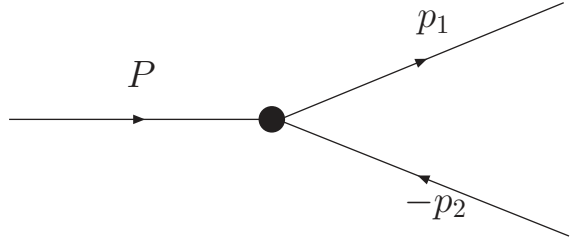}}
\end{tabular}
\caption{The vertex for meson-quark-antiquark.}\label{fig:LFQM1}
\end{figure}
\end{center}

In the momentum representation, the wavefunction  $\Psi^{JJ_z}_{LS}$ for the state
$^{2S+1}L_{J}$ can be decomposed into the form
\begin{eqnarray}\label{vf4}
\Psi^{JJ_z}_{LS}(\tilde{p_1},\tilde{p_2},\lambda_1,\lambda_2)=\frac{1}{\sqrt{N_c}}\langle
L S;L_z S_z|L S; J J_z\rangle
R^{SS_z}_{\lambda_1\lambda_2}(x,p_\perp)\varphi_{LL_z}(x,p_\perp),
\end{eqnarray}
where $\varphi_{LL_z}(x,p_\perp)$ describes the relative momentum
distribution of the quark (antiquark)
in the meson and $L$ is the orbital angular momentum between the constituents.
$R^{SS_z}_{\lambda_1\lambda_2}$ transforms  a light-front helicity
 $(\lambda_1,\lambda_2)$ eigenstate  to a state with
definite spin $(S,S_z)$  and it is expressed as
\begin{eqnarray}\label{vf5}
R^{SS_z}_{\lambda_1\lambda_2}(x,p_\perp)&&=\frac{1}{\sqrt{2}\tilde{M_0}(M_0+m_1+m_2)}\bar{u}(p_1,\lambda_1)(\bar{\slash\!\!\!\!
P}+M_0)\Gamma_S v(p_2,\lambda_2),\nonumber\\
\Gamma_0&&=\gamma_5\,\,\,\,\,\,\,\,(\,{\rm for}\, S=0),\nonumber\\
\Gamma_1&&=-\slash\!\!\!\hat{\varepsilon}(S_z)\,\,\,\,\,\,\,\,(\,{\rm
for}\, S=1),
\end{eqnarray}
with $\bar P=p_1+p_2$,
$M_0^2=\frac{m_1^2+p_\perp^2}{x_1}+\frac{m_2^2+p_\perp^2}{x_2}$
and $\tilde{M_0}\equiv\sqrt{M_0^2-(m_1-m_2)^2}$. For more details, readers are suggested to refer
Ref.\cite{Cheng:2003sm}.

In the LFQM,   the harmonic oscillator wavefunctions are employed
to describe the relative 3-momentum distribution of quark and
antiquark in a meson. For d-wave  the harmonic oscillator
wavefunction is\cite{Faiman:1968js}
\begin{eqnarray}\label{vf6}
\varphi_{_{2 L_z}}(x,p_\perp)=\hat{\varepsilon}^{\mu\nu}(L_z)K_\mu
K_\nu\frac{\sqrt{2}}{\beta^2}\varphi,
\end{eqnarray}
where $K=(p_2-p_1)/2$ and $\varphi$ is the harmonic oscillator
wavefunction for s wave and its explicit expression is with
 \begin{eqnarray}\label{app2}
&&\varphi=4(\frac{\pi}{\beta^2})^{3/4}\sqrt{\frac{dp_z}{dx_2}}{\rm
 exp}(-\frac{p^2_z+p^2_\perp}{2\beta^2}),\nonumber\\&&
p_z=\frac{x_2M_0}{2}-\frac{m_2^2+p^2_\perp}{2x_2M_0}.
 \end{eqnarray}

Substituting  Eq.(\ref{vf4}) into Eq.(\ref{vf3}), we  deduce the
expression
\begin{eqnarray}\label{vf7}
\Psi^{JJ_z}_{2S}(\tilde{p_1},\tilde{p_2},\lambda_1,\lambda_2)&&=\frac{1}{\sqrt{N_c}}\langle
2 S;L_z S_z|2 S; J J_z\rangle
R^{SS_z}_{\lambda_1\lambda_2}(x,p_\perp)\varphi_{_{2L_z}}(x,p_\perp)\nonumber\\&&
=\frac{1}{\sqrt{N_c}}\frac{\varphi}{\sqrt{2}\tilde{M}_0(M_0+m_1+m_2)}\bar{u}(p_1,\lambda_1)(\bar{\slash\!\!\!\!
P}+M_0)\Gamma_{(^{2s+1}D_J)}v(p_2,\lambda_2),
\end{eqnarray}
with
\begin{eqnarray}\label{vf8}
&&\Gamma_{(^{3}D_1)}=\sqrt{\frac{6}{5\beta^4}}[-\frac{\slash\!\!\!\hat{\varepsilon}(J_z)}{3}(\frac{K\cdot
\bar{P}^2}{M_0^2}-K^2)
+\frac{\slash\!\!\!\!\bar{P} K\cdot \bar{P}K\cdot\hat{\varepsilon}(J_z)}{M_0^2}-\slash\!\!\!\!{K} K\cdot\hat{\varepsilon}(J_z)],\nonumber\\
&&\Gamma_{(^{1}D_2)}=\sqrt{\frac{2}{\beta^4}}\hat{\varepsilon}^{\mu\nu}(J_z)K_\mu K_\nu\gamma_5,\nonumber\\
&&\Gamma_{(^{3}D_2)}=\sqrt{\frac{4}{3\beta^4}}\hat{\varepsilon}^{\mu\nu}(J_z)\gamma_5[\gamma_\nu\gamma_\mu\frac{(K\cdot\bar
P)^2-M_0^2K^2}{3M_0}-\gamma_\mu K_\nu(K\cdot\bar P-M_0\slash
\!\!\!\!K)+K_\mu K_\nu],\nonumber\\
&&\Gamma_{(^{3}D_3)}=\sqrt{\frac{2}{9\beta^4}}\hat{\varepsilon}^{\mu\nu\alpha}(J_z)\gamma_\beta
 (K_\mu K_\nu g_{\alpha\beta}+K_\mu K_\alpha g_{\nu\beta}+K_\alpha K_\nu g_{\mu\beta}),
\end{eqnarray}
where relations $\langle 2\, 0;L_z\, 0|2\, 0; 2\,
J_z\rangle=\hat{\varepsilon}^{*\mu\nu}(L_z)\hat{\varepsilon}_{\mu\nu}(J_z)$,
$\langle 2\, 1;L_z\, S_z|2\, 1; 1\,
J_z\rangle=-\sqrt{\frac{3}{5}}\hat{\varepsilon}^{*\mu\nu}(L_z)\hat{\varepsilon}_{*\mu}(S_z)\hat{\varepsilon}_{\nu}(J_z)$
, $\langle 2\, 1;L_z\, S_z|2\, 1; 2\,
J_z\rangle=i\sqrt{\frac{2}{3}}\epsilon^{\alpha\beta\mu\nu}\hat{\varepsilon}^{*}_{\alpha\omega}(L_z)\hat{\varepsilon}^*_{\beta}(S_z)
\hat{\varepsilon}_{\mu\omega'}(J_z)g^{\omega\omega'}\frac{\bar{P}_\nu}{M_0}$
and $\langle 2\, 1;L_z\, S_z|2\, 1; 3\,
J_z\rangle=\frac{1}{3}\hat{\varepsilon}^{\mu\nu\alpha}(J_z)[\hat{\varepsilon^*}_{\mu\nu}(L_z)\hat{\varepsilon^*}_{\alpha}(S_z)
+\hat{\varepsilon^*}_{\mu\alpha}(L_z)\hat{\varepsilon^*}_{\nu}(S_z)+\hat{\varepsilon^*}_{\nu\alpha}(L_z)\hat{\varepsilon^*}_{\mu}(S_z)]$
are used.

One can further simplify these wavefunctions in terms of the Dirac
equation $\slash\!\!\!p_1 u(p_1)=m_1u(p_1)$ and $\slash\!\!\!p_2
v(p_1)=-m_2v(p_1)$, so that all scalar products of vectors are
replaced by  only $M_0$, $m_1$ and $m_2$ via a simple algebra, thus
the wave function is
\begin{eqnarray}\label{vf71}
\Psi^{JJ_z}_{2S}(\tilde{p_1},\tilde{p_2},\lambda_1,\lambda_2)
=\bar{u}(p_1,\lambda_1)h'_{(^{2s+1}D_J)}\Gamma'_{(^{2s+1}D_J)}v(p_2,\lambda_2),
\end{eqnarray}
where
\begin{eqnarray}\label{vf101}
&&h'_{(^{3}D_1)}=-\sqrt{\frac{1}{N_c}}\frac{1}{\sqrt{2}\tilde{M_0}}
\frac{\sqrt{6}}{12\sqrt{5}M_0^2\beta^2}[M_0^2-(m_1-m_2)^2][M_0^2-(m_1+m_2)^2]\varphi,\nonumber\\
&&h'_{(^{1}D_2)}=\sqrt{\frac{1}{N_c}}\frac{1}{\tilde{M_0}\beta^2}\varphi,\nonumber\\
&&h'_{(^{3}D_2)}=\sqrt{\frac{1}{N_c}}\sqrt{\frac{2}{3}}\frac{1}{\tilde{M}_0\beta^2}\varphi,
\nonumber\\
&&h'_{(^{3}D_3)}=\sqrt{\frac{1}{N_c}}\frac{1}{3}\frac{1}{\tilde{M}_0\beta^2}\varphi,
\end{eqnarray}

and
\begin{eqnarray}\label{vf91}
&&\Gamma'_{(^{3}D_1)}=[\gamma_\mu-\frac{1}{w_{(^{3}D_1})}(p_1-p_2)_\mu]\hat{\varepsilon}^{\mu},\nonumber\\
&&\Gamma'_{(^{1}D_2)}=\gamma_5K_\mu K_\nu\hat{\varepsilon}^{\mu\nu},\nonumber\\
&&\Gamma'_{(^{3}D_2)}=\gamma_5[\frac{1}{w^a_{(^{3}D_2)}}\gamma_\nu\gamma_\mu+\frac{1}{w^b_{(^{3}D_2)}}\gamma_\mu
K_\nu+\frac{1}{w^c_{(^{3}D_2)}}\gamma_\nu
K_\mu+\frac{1}{w^d_{(^{3}D_2)}}K_\mu
K_\nu]\hat{\varepsilon}^{\mu\nu},\nonumber\\
&&\Gamma'_{(^{3}D_3)}=[K_\mu
K_\nu(\gamma_\alpha+\frac{2K_\alpha}{w_{(^{3}D_3)}})+K_\mu
K_\alpha(\gamma_\nu+\frac{2K_\nu}{w_{(^{3}D_3)}})+K_\alpha
K_\nu(\gamma_\mu+\frac{2K_\mu}{w_{(^{3}D_3)}})]\hat{\varepsilon}^{\mu\nu\alpha},
\end{eqnarray}
with
\begin{eqnarray}\label{vf92}
&&w_{(^{3}D_1)}=\frac{(m_1+m_2)^2-M_0^2}{2M_0+m_1+m_2},\nonumber\\
&&w^a_{(^{3}D_2)}=\frac{-12M_0^2}{[M_0^2-(m_1+m_2)^2][M_0^2-(m_1-m_2)^2]},\nonumber\\
&&w^b_{(^{3}D_2)}=-\frac{6M_0^2}{(2M_0+m_1+m_2)[M_0^2-(m_1-m_2)^2]},\nonumber\\
&&w^c_{(^{3}D_2)}=\frac{6M_0^2}{(M_0-m_1-m_2)[M_0^2-(m_1-m_2)^2]},\nonumber\\
&&w^b_{(^{3}D_2)}=\frac{M_0}{m_2-m_1},\nonumber\\
&&w_{(^{3}D_3)}=M_0+m_1+m_2.
\end{eqnarray}

\begin{center}
\begin{figure}[htb]
\begin{tabular}{c}
\scalebox{0.8}{\includegraphics{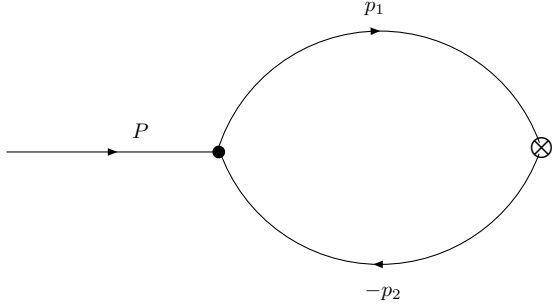}}
\end{tabular}
\caption{The Feynman diagram for a meson annihilation.}\label{DM2}
\end{figure}
\end{center}

{It is interesting to ask where the QCD which definitely governs
the physical processes, gets involved or how can one implant the
QCD information into our calculation in the LFQM. In
Ref.\cite{Glazek:2006cu} the authors derived an effective
Hamiltonian for bound states in the light-front frame based on the
standard Lagrangian of  QCD. The vertex function is the effective
coupling between the bound state and its constituent quarks, thus
as the wavefunction of the bound state is obtained the effective
vertex function is in hand. After a long discussion about the
Lorentz structure and the features of the dynamics of the vertex
functions, they derive the vertex function which has exactly the
form of Eq.(\ref{vf71}) in our work. In Ref.\cite{Glazek:2006cu},
the radial wave function  was obtained by solving the eigenvalue
equation numerically. Obviously all QCD information (both
short-distance and long-distance effects) is involved in the
Hamiltonian and as well as in the solution. As argued in
literature\cite{Faiman:1968js,Cheng:1996if}, the solution can be
well approximated by a Gaussian function with model parameters to
be fixed by fitting data. Thus following Ref.\cite{Cheng:2003sm}
we choose a Gaussian wave function where the QCD information is
included in the model parameter $\beta$.}

We can apply these wavefunctions to deal with concrete physical
processes. For example, when we calculate the rate of $^3D_1$
state annihilation through a vector current (in Fig.\ref{DM2}),
the transition amplitude is written as
\begin{eqnarray}\label{vf9.0}
\mathcal{A}_\mu^{conv}=&&N_c\int d^3\tilde{
p_1}\Psi_{2\,1}^{1\,J_z}\frac{\bar
v(p_2,\lambda_2)}{\sqrt{p_2^+}}\gamma_\mu\frac{u(p_1,\lambda_1)}{\sqrt{p_1^+}}\nonumber\\=&&
N_c\int\frac{ dx_1 d^2p_\perp}{16\pi^3}
\frac{h'_{^3D_1}}{\sqrt{x_1x_2}}{\rm
Tr}[\Gamma'_{^3D_1}(\slash\!\!\!\!p_2-m_2)\gamma_\mu(\slash\!\!\!\!p_1+m_1)]
\nonumber\\=&&N_c\int\frac{ dx_1 d^2p_\perp}{16\pi^3}
\frac{h'_{^3D_1}}{\sqrt{x_1x_2}}{\rm
Tr}\{[\gamma_\nu-\frac{(p_1-p_2)_\nu}
{w_{^3D_1}}](\slash\!\!\!\!p_2-m_2)\gamma_\mu(\slash\!\!\!\!p_1+m_1)\}\hat{\varepsilon}^\nu.
\end{eqnarray}

\section{Vertex functions in the covariant light-front approach}

Comparing with the conventional LFQM where both $p_1$ and $p_2$
are on their mass shells, in the covariant light-front approach
the quark and antiquark are off-shell, but the total momentum
$P=p_1+p_2$ is the on-shell momentum of the meson, i.e. $P^2=M^2$
where $M$ is the mass of the meson. Obviously, the covariant LFQM
is closer to the physical reality.

{ If one tries to obtain the covariant vertex functions based on an
underlying principle, i.e. QCD, he should invoke a reasonable
theoretical framework. To directly obtain the covariant vertex
functions in the 4-dimensional momentum space, the authors of
Ref.\cite{Jaus:1999zv} suggested to solve the Bethe-Salpeter (B-S)
equations for the bound states \cite{BS,Roberts:1994dr}. The kernel
of the B-S equation includes the Coulomb piece which is induced by
the one-gluon exchange as well as its higher-order corrections, and
the confinement piece which incorporates the non-perturbative QCD
but is not derivable so far. Genrally for solving the B-S equation,
the instantaneous approximation is usually taken. }

{ In the concrete calculations of the observable physical quantity
in terms of the LFQM the final result is eventually reduced into an
integration over the four-momentum. Fortunately, by doing so, we may
not really need the explicit covariant wavefunctions defined in the
four-momentum space. Namely, we try to reduce the integration into a
simple form where only three-momentum wavefunctions remain by a
mathematical manipulation, then we are able to relate the
corresponding integrand to the conventional vertex function which is
well defined in the three-momentum space. }

Since the Lorentz structures of the covariant vertices are the
same as that of the conventional vertex functions we rewrite these
covariant vertex functions  in Eq.(\ref{vf71}) as
\begin{eqnarray}\label{vf9}
&&iH_{(^{3}D_1)}[\gamma_\mu-\frac{1}{W_{(^{3}D_1)}}(p_1-p_2)_\mu]{\varepsilon}^{\mu},\nonumber\\
&&iH_{(^{1}D_2)}\gamma_5K^\mu K^\nu{\varepsilon}^{\mu\nu},\nonumber\\
&&iH_{(^{3}D_2)}\gamma_5[\frac{1}{W^a_{(^{3}D_2)}}\gamma^\omega\gamma^\mu+\frac{1}{W^b_{(^{3}D_2)}}\gamma^\mu
K^\omega+\frac{1}{W^c_{(^{3}D_2)}}\gamma^\omega
K^\mu+\frac{1}{W^d_{(^{3}D_2)}}K^\mu
K^\omega]{\varepsilon}^{\mu\nu},
\nonumber\\
&&iH_{(^{3}D_3)} [K_\mu
K_\nu(\gamma_\alpha+\frac{2K_\alpha}{W_{(^{3}D_3)}})+K_\mu
K_\alpha(\gamma_\nu+\frac{2K_\nu}{W_{(^{3}D_3)}})+K_\alpha
K_\nu(\gamma_\mu+\frac{2K_\mu}{W_{(^{3}D_3)}})]{\varepsilon}^{\mu\nu\alpha},
\end{eqnarray}
where $H_{(^{2S+1}D_J)}$ and $W_{(^{2S+1}D_J)}$ are functions in
the 4-dimensional space.  Practically, the vertex function(s)
is(are) included in a transition matrix element, for example, the
amplitude of $^3D_1$ state annihilation via a vector current is
written as
\begin{eqnarray}\label{vf9.1}
\mathcal{A}_\mu^{cov}&&=-i^2\frac{N_c}{16\pi^4}\int d^4p_1
\frac{H_{(^{3}D_1)}}{N_1N_2}{\rm
Tr}\{[\gamma_\nu-\frac{(p_1-p_2)_\nu}
{W_{^3D_1}}](-\slash\!\!\!p_2+m_2)\gamma_\mu(\slash\!\!\!p_1+m_1)\}{\varepsilon}^\nu\nonumber\\
&&=\frac{-i^2N_c}{16\pi^4}\int d^4p_1
\frac{H_{(^{3}D_1)}}{N_1N_2}\,s\,{\varepsilon}^\nu,
\end{eqnarray}
where $s={\rm Tr}\{[\gamma_\nu-\frac{(p_1-p_2)_\nu}
{W_{^3D_1}}](-\slash\!\!\!p_2+m_2)\gamma_\mu(\slash\!\!\!p_1+m_1)\}$,
$N_1=p_1^2-m_1^2+i\epsilon$ and $N_2=p_2^2-m_2^2+i\epsilon$. One
first needs to integrate over $p_1^-$ as discussed in
Ref.\cite{Jaus:1999zv,Cheng:2003sm}. Integrating over $p_1^-$ is
completed by a contour integration where the antiquark is set on
shell. Then the integration turns into
\begin{eqnarray}\label{vf9.2}
\frac{N_c}{16\pi^3}\int dx_1d^2p_\perp\frac{h_{(^{3}D_1)}}{x_2
x_1(M^2-M_0^2)}\hat s\,\hat{\varepsilon}^\nu,
\end{eqnarray}
where $w_{(^{3}D_1)}$ and $\hat{\varepsilon}^{\nu}$  replace
$W_{(^{3}D_1)}$ and ${\varepsilon}^{\nu}$ in Eq.(\ref{vf9.1})
respectively.

Following
Ref.\cite{Cheng:2003sm} we have the relation
\begin{eqnarray}\label{vf10.0}
h_{(^{3}D_1)}=(M^2-M_0^2)\sqrt{x_1x_2}h'_{(^{3}D_1)}.
\end{eqnarray}
An additional factor $(M^2-M_0^2)\sqrt{x_1x_2}$  was
introduced when comparing the decay constant $f_P$ of pseudoscalar meson obtained in
the two approaches as depicted in the appendix A of Ref.\cite{Cheng:2003sm}.
The legitimacy is guaranteed because the decay constant is free of zero mode contribution. Then the authors have applied the relation into the
vertex functions for S and P waves.
To show the reasonability of such replacement, we substitute Eq.(\ref{vf10.0}) into
Eq.(\ref{vf9.2}) to obtain a new expression whose form is similar to the right side of
Eq.(\ref{vf9.0}). However, the trace in Eq. (\ref{vf9.2})  involves the zero mode contribution which makes its form different from that in Eq.(\ref{vf9.0}).
 Generally, after the contour integration over
$p_1^-$, $h_{(^{2S+1}D_J)}$, $w_{(^{2S+1}D_J)}$ and
$\hat{\varepsilon}$ replace $H_{(^{2S+1}D_J)}$, $W_{(^{2S+1}D_J)}$
and ${\varepsilon}$ respectively with the following relation to
the corresponding quantities of the conventional LFQM
\begin{eqnarray}\label{vf10}
h_{(^{2S+1}D_J)}=(M^2-M_0^2)\sqrt{x_1x_2}h'_{(^{2S+1}D_J)}.
\end{eqnarray}

{Here we only concern the form of the covariant vertex function
for the D-wave, including its Lorentz structure and coefficient,
as well as its relations to the conventional vertex function. The
details about the S- and P wave vertex functions were discussed in
earlier literature\cite{Cheng:2003sm}. When one needs to calculate
a transition matrix in the covariant light-front quark model, he
must know those vertex functions. Jaus has analyzed  the case of
the covariance of the transition matrix \cite{Jaus:1999zv}, and in
his work, a general form of vertex function is used and the
three-momentum conservation is automatically guaranteed. He
\cite{Jaus:1999zv} indicates that the general form of the vertex
function $h$ must be functions of $\hat{N_i}=x_i(M^2-M^2_0)\;
(i=1,2)$. Obviously the function $h$ adopted in this work
coincides with this requirement.}
\section{The formula for $^3D_1$ state}
For a $J^{PC}=1^{--}$  state, the the orbital momentum between the
two constituents may be $L=0$ (s-wave) or $L=2$ (d-wave) and their
total spin is 1 ($S=1$). In Ref.\cite{Cheng:2003sm} the authors
gave the meson-quark-antiquark vertex  for $^3S_1$ state as
\begin{eqnarray}\label{vf11}
iH_V[\gamma_\mu-\frac{1}{W_{V}}(p_1-p_2)_\mu].
\end{eqnarray}
Carrying out the contour
integration over $p_1^-$, $H_V$ and $W_V$ turn into $h_V$ and $w_V$
$$h_V=(M^2-M_0^2)\sqrt{\frac{x_1x_2}{N_c}}\frac{1}{\sqrt{2}\tilde{M}_0}\varphi,$$
$$w_V=M_0+m_1+m_2,$$
where  the subscript $V$ only refers to $^3S_1$ state.

The Lorentz structure of the vertex functions for $^3D_1$
and $^3S_1$ states are the same
because they have the same quantum number $J^{PC}$. The difference between the s-wave and
d-wave is included in the coefficient functions $h_M$ and $w_M$.

The decay amplitude of an $^3S_1$ state via a vector current is proportional to \cite{Cheng:2003sm}
\begin{eqnarray}\label{vf10.1}
A_\mu=-i^2\frac{N_c}{(2\pi)^4}\int
d^4p_1\frac{iH_{V}}{N_1N_2}Tr\{\gamma_\mu(\slash\!\!\!p_1+m_1)[\gamma_\nu-\frac{(p_1-p_2)_\nu}
{W_{V}}](-\slash\!\!\!p_2+m_2)\}\hat{\varepsilon}^\nu,
\end{eqnarray}
which is the same as Eq.(\ref{vf9.1}), except  $H_{^3D_1}$ and
$W_{^3D_1}$ are replaced by $H_V$ and $W_V$ respectively.
Integrating over $p_1^-$ $H_{^3D_1}$, $W_{^3D_1}$, $H_V$ and $W_V$
reduce into $h_{^3D_1}$, $w_{^3D_1}$, $h_V$ and $w_V$. The decay
constant for $^3S_1$ state reads
\begin{eqnarray}\label{vf11}
f_V=\frac{N_c}{4\pi^3M}\int
d{x_2}d^2p_\perp\frac{h_V}{x_1x_2(M^2-M_0^2)}[x_1M_0^2-m_1(m_1-m_2)-p_\perp^2+\frac{m_1+m_2}{w_V}p_\perp^2]
,
\end{eqnarray}
so that one would obtain the decay constant of the $^3D_1$ state by  replacing
$h_V$ and $w_V$ by $h_{^3D_1}$ and $w_{^3D_1}$, thus it is
\begin{eqnarray}\label{vf12}
f_{^3D_1}=\frac{N_c}{4\pi^3M}\int
d{x_2}d^2p_\perp\frac{h_{^3D_1}}{x_1x_2(M^2-M_0^2)}[x_1M_0^2-m_1(m_1-m_2)-p_\perp^2+\frac{m_1+m_2}{w_{^3D_1}}p_\perp^2]
.
\end{eqnarray}

In fact since the Lorentz structure of the vertex functions for
$^3D_1$ and $^3S_1$ are the same all the formula for $^3D_1$ can
be deduced from those for $^3S_1$. For example, the form factors
$f, g, a_+$ and $a_-$ of $P\to ^3D_1(V)$ decay  can be obtained by
simply replacing $h_V$ and $w_V$ of $f, g, a_+$ and $a_-$ given in
Ref.\cite{Jaus:1999zv,Cheng:2003sm}.

With these formula  we will be able to explore some new resonances of angular excited states, or furthermore to study
the mixing of $2S-1D$ which was proposed to explain the famous $\rho-\pi$ puzzle for
$\psi'$\cite{Ding:1991vu,Rosner:2001nm,Kuang:1989ub} in this
model.

\section{A brief summary}
In this paper we  deduce the vertex functions (or wave functions)
for the d-wave in the conventional and covariant light-front quark
model.

For the $^3D_1$ state the $J^{PC}$ is $1^{--}$ and the Lorentz
structure of its wave function is the same as that for the $^3S_1$
state so we obtain some useful formula for $^3D_1$ from the
formula for $^3S_1$ given in Ref.\cite{Jaus:1999zv,Cheng:2003sm}.
It is noted we just discuss the vertex functions ($i\Gamma_M$) for
the incoming meson whereas for the outgoing meson the
corresponding vertex functions should be
$i(\gamma_0\Gamma^\dagger_M\gamma_0)$\cite{Cheng:2003sm}.

Since we adopt the Gaussian-type function for the radial part of the whole wavefunction instead of
a solution obtained by solving the Schr\"odinger equation or the B-S equation,  the simplification definitely brings up certain theoretical
uncertainties, but as more data will be collected in the future,
the more precise model parameter(s) will be determined and even
the form of the wavefunction can be improved, thus we may do a
better job along the line.

These vertex functions can be employed when one calculates
the transition rates in this model. In the future we will study some concrete physical
transitions where d-wave mesons are involved in terms of these vertex functions. The results will be compared with data
and the consistency would tell us the accuracy degree of the model and the derived vertex functions. Once
the validity of the model is verified via some processes, we can further discuss some long-standing puzzles
and help to identify new resonances which are continuously observed at BES and BELLE and elsewhere.

\section*{Acknowledgments}

We thank Dr. Y. Chen for introducing new developments and progresses in
the lattice QCD theory. This work is supported by the National Natural Science Foundation
of China (NNSFC) under the contract No. 11075079 and No. 11005079;
the Special Grant for the Ph.D. program of Ministry of Eduction of
P.R. China No. 20100032120065.

\end{document}